\documentstyle[12pt]{article}

\newcommand{\lr}{\leftrightarrow}

\newcommand{\up}{\uparrow}

\newcommand{\dn}{\downarrow} 

\newcommand {\dis}{\displaystyle}

\input epsf

\begin{document}

\begin{center}
{\Large\bf The Origin of Degeneracies and Crossings in the 1d Hubbard Model}
 
\vspace{1cm}

Emil~A.~Yuzbashyan$^1$, Boris~L.~Altshuler$^{1,2}$ and 
B.~Sriram~Shastry$^{3, 4}$

\vspace{0.5cm}

{\small\it

\noindent $^1$ Physics Department, Princeton University, Princeton, NJ 08544\\
 $^2$ NEC Research Institute, 4 Independence Way, 
Princeton, NJ 08540\\
 $^3$  Bell Laboratories Lucent
Technologies, 600 Murray Hill, NJ 07974\\
 $^4$ Indian Institute of Science, Bangalore 560012, India\\

}
 
\end{center}
\begin{abstract}
The paper is devoted to the connection between integrability 
of a finite quantum system and degeneracies of its energy levels.
In particular, we analyze in detail the energy spectra 
of finite Hubbard chains. 
Heilmann and Lieb demonstrated that in these systems there are crossings
of levels of
the same parameter independent symmetry. 
We show that this apparent violation of the Wigner-von Neumann 
noncrossing rule follows directly  from the 
existence of nontrivial conservation
laws and is a characteristic signature of quantum integrability.
The energy spectra of Hubbard chains display many instances of permanent (at
all values of the coupling) twofold degeneracies that cannot be explained
by parameter independent symmetries. We relate these degeneracies to
the different transformation  
properties of the conserved currents under spatial reflections and
the particle-hole transformation and  estimate the fraction of
doubly degenerate states.  We also discuss multiply
degenerate eigenstates of the Hubbard Hamiltonian.  The wave functions
of many of these states do not depend on the coupling, which suggests the
existence of an additional parameter independent symmetry.  
\end{abstract}

\section{Introduction}
 The close connection between symmetry and degeneracy has been explored since
the foundation of Quantum Mechanics.   Famous examples include  degeneracies 
of spectra in angular momentum in  the Hydrogen
atom (well known as the accidental degeneracy) \cite{Pauli} and
the 3d harmonic oscillator \cite{LL}. In the present paper we focus on 
the  implications  of the rich symmetry
structure for the spectrum of the 1d Hubbard model as well as
for a general class of quantum integrable systems. We analyze in detail how
degeneracies and other spectral properties of the Hubbard Hamiltonian 
reflect the symmetries of the model.
  
Here we restrict ourselves to the case of Hamiltonians that 
depend on a single real parameter, referred to as a coupling. 
In this case one can distinguish between
two types of degeneracy.
 The first type, often
called permanent degeneracy,  
 refers to energy levels which remain degenerate for all
values of the parameter. 
 The appearance of permanent  degeneracies  suggests the  existence of
non--commuting symmetry operators (see e.g. \cite{LL}).
    
The second type
of degeneracy is a crossing of energy levels  that occurs  at 
a particular value of the coupling. Multiple level crossings at a certain
point indicate a higher symmetry of the system at this particular point as
compared to  other points.
 For example, the limits of zero or infinite coupling frequently
have  enhanced symmetry. 

Simple pairwise crossings  are also interesting from a different
perspective. As is well known, 
in the absence of symmetries energy levels repel, or, equivalently,
crossings of levels   are prohibited  unless the states have different 
symmetry. This
{\it noncrossing rule} was suggested by Hund in 1927 \cite{Hund},
and justified by  von Neumann and Wigner \cite{NW} two years 
latter. 
Since then the level repulsion phenomenon has been revisited and
elaborated upon in various contexts, see for example 
\cite{Beenakker, Berry, Arnold}
 and references therein. 
Usually textbooks, e.g., \cite{LL, M, Davydov}, present a
 simplified version of the justification due to Teller \cite{Teller}. 

\begin{figure}[ht]
\epsfysize=3.0in 
\epsfxsize=4.0in  
\centerline{\epsfbox{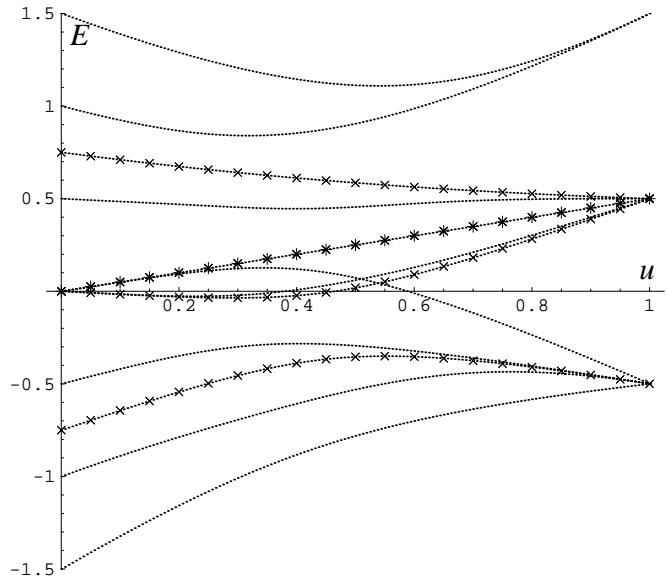}}
\caption{ Energies for 
$ (P, \sigma,   I^{(o)}, S, L )=
(3, 1, 1, 0, 1 ) $  in units of $U-4T$ as functions
of $u=U/(U-4T)$. Levels
marked with crosses are twofold degenerate. They also correspond to 
$ (P, \sigma, I^{(o)}, S, L )=(3, -1, 1, 0, 1 )$. Upside down the
figure shows levels for the same set of quantum numbers with 
$S \leftrightarrow L$. ***** shows the seven fold degenerate level. }
\label{h1}
\end{figure}

The mathematical validity of the 
noncrossing rule
depends crucially on the interpretation of the word symmetry, 
normally understood as a space (or internal space) symmetry.  
 This was highlighted by an interesting example
by Heilmann and Lieb in 1971 \cite{HL}. These authors
 pointed out  that the noncrossing rule is
apparently violated in the case of the  1d Hubbard Hamiltonian for 
the benzene
 molecule
if only those  symmetries that do not depend on the coupling constant
are taken into account. 
Indeed, Fig.~\ref{h1} 
 illustrates  that
  a substantial number of crossings  can be found even 
within the subsets of levels characterized by the same set of quantum numbers.
 Based on this fact, Heilmann and Lieb 
concluded that ``there must be a natural parameter dependent 
group''to account for these violations. 

The 1d Hubbard model is solvable by Bethe's {\it Ansatz}, and usually
the applicability of this {\it Ansatz} is understood to rest upon 
{\it Quantum Integrability} of the model,
i.e the existence of an infinite  number of mutually   commuting operators. 
Such commuting operators are often alternately
termed as  ``conserved currents '', or ``dynamical conservation laws'',
 or simply ``dynamical symmetries'', and 
 are invariably associated with all known quantum integrable models.
However, it was not until 1986 when the
 parameter dependent integrals of motion (conserved currents),
 were identified by Shastry \cite{S1, S2, S3}, who constructed
 a transfer matrix 
that commuted with the Hubbard Hamiltonian.

The Hubbard Hamiltonian to be considered
consists of the kinetic term, which allows electrons to hop between 
the nearest neighbors on a regular polygon (i.e. periodic
boundary conditions are assumed) and the 
Coulomb interaction between electrons of opposite spin on the same site.
\begin{equation}
\hat H=T\sum_{j=1}^N\sum_{s=\up\dn}
(c^\dagger_{js}c_{j+1\,s}+c^\dagger_{j+1\,s}c_{js})+ 
U\sum_{j=1}^N\left(\hat n_{j\uparrow}-
\frac{1}{2}\right)\left(\hat n_{j\downarrow}-
\frac{1}{2}\right)
\label{H}
\end{equation}
where $c^\dagger_{js}$ and  $c_{js}$ are the creation and 
annihilation operators for an electron of spin projection $s$ on site $j$ and 
$\hat n_{js}=c^\dagger_{js}c_{js}$ is the number operator.  The goal of this 
paper 
is to
study how the interplay between parameter dependent integrals of motion and
ordinary (parameter independent) symmetries is manifested in 
permanent degeneracies
and level crossings in the energy spectrum. The Hamiltonian
(\ref{H}) is chosen as 
a tutorial example of a many body system that has both parameter
dependent and parameter independent symmetries.

The paper is organized as follows. Section~2 is devoted to parameter
independent symmetries of the Hubbard model that are later used in Section~4
to diagonalize the Hamiltonian and assign symmetry quantum numbers
such as the total momentum, spin, particle--hole symmetry etc. to all
states.  
  Section~3 discusses
the structure and properties of parameter dependent conservation laws.
 Permanent degeneracies are treated in detail in Section~5.
  In the same section we
consider  the
multiple permanent degeneracies  and mention
spectral properties of the dynamical conservation laws. 

In Section 6 we formulate
the noncrossing rule in the framework of adiabatic equations of motion for
matrix elements of the Hamiltonian and higher currents. This language is
useful for understanding the suppression of level repulsion
in integrable models. Next, based on numerical analysis, we discuss
a curious behavior of the transverse matrix element in the vicinity of
crossings specific to the Hubbard Hamiltonian.
Finally, to illustrate the connection between level crossings and integrability
we consider in some detail a simple case of integrable $3\times 3$ matrix
systems (Section 6). In this case we conclude that there are no 
``accidental degeneracies'', i.e. all 
degeneracies
may be associated with dynamical conservation laws.

The majority of the results do not depend on the number of
sites ($N$) or electrons
 ($M$). However, cases of even and odd $N$ and $M$
have to be treated
differently, the latter case being substantially simpler 
because of less symmetry that
needs to be taken into account. We therefore take for concreteness that both 
these numbers are
even, $N=2n$ and $M=2m$ with an equal number of spin up and spin down 
electrons. Throughout the paper analytical results are
illustrated by numerical computations for a special case of 
benzene ($m=n=3$), and a few other choices of $(m, n)$.
 
\section{U--independent Symmetries}

The starting point of our analysis of the spectral properties of the
Hubbard Hamiltonian (\ref{H}) is the list of its
U--independent symmetries.
 We will follow closely  
the paper by Heilmann and Lieb \cite{HL} adopting in most cases
 their notation.  

The role of the U--independent symmetries is twofold.
First, they greatly simplify  the diagonalization of the Hamiltonian and
integrals of motion and provide a convenient labeling for various parts of
the spectrum. Besides, for the purposes of studying specifically 
the effect
of parameter dependent conservation laws all these symmetries have to be 
factored out  anyway.
 
\begin{figure}[ht]
 \epsfxsize=6cm
\centerline{\epsfbox{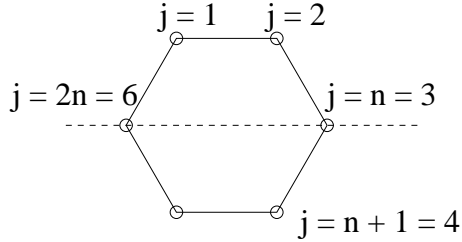}}
\caption{Regular hexagon ($n=3$). $\hat C_6$ rotates by $\pi/3$. 
$\hat\sigma$ reflects in a
line through vertices 3 and 6.}
\label{poly}
\end{figure}

U--independent symmetries fall into three
major categories -- the symmetry of the polygon, the spin symmetry, and
the particle-hole symmetry. Spatial symmetries can be generated by two
operators: operator $\hat C_{2n}$  that rotates the polygon by $\pi/n$, and 
operator $\hat\sigma$ that
reflects it in a line through vertices $n$ and $2n$ (see Fig.~\ref{poly}). 
These operators can be expressed in 
terms
of on site creation and annihilation operators as
\begin{equation}
\hat C_{2n}=\hat\sigma\hat\sigma'\quad\hat\sigma=
\prod_{s=\up,\dn}\prod_{k=1}^{n-1}{\bf\hat J}_{ks;-ks}\quad
\hat\sigma'=\prod_{s=\up,\dn}\prod_{k=1}^n {\bf\hat J}_{k-1,s; 2n-k,s}
\label{space}
\end{equation}
where ${\bf\hat J}_{js,j's'}$ interchanges the orbitals $(js)$ and $(j's')$
\begin{equation}
{\bf \hat J}_{js,j's'}=1-\hat n_{js}-\hat n_{j's'}+c^\dagger_{js}c_{j's'}+
c^\dagger_{j's'}c_{js}
\label{orbitals}
\end{equation}
Eigenvalues of $\hat C_{2n}$ are $e^{i\pi P/n}$, where $P$ is an integer 
$0\le P\le 2n-1$ that represents the total momentum of the state. 
The reflection operator $\hat\sigma$ has  eigenvalues $\pm1$.

The spin symmetry can be generated by ${\bf\hat S}^2$
and $\hat S_z$
\begin{equation}
\hat S_z=\frac{{\bf\hat n}_\up-{\bf\hat n}_\dn}{2}\quad 
\hat S_+=(\hat S_-)^\dagger=\sum_{j=1}^{2n}c^\dagger_{j\up}c_{j\dn}
\label{spinsu2}
\end{equation}
\begin{equation}
{\bf\hat S}^2=\frac{\hat S_-\hat S_++\hat S_+\hat S_-}{2}+\hat S_z^2
\label{s2} 
\end{equation}
where
\begin{equation}
 {\bf\hat n}_s=\sum_{j=1}^{2n} \hat n_{js} 
 \label{n}
\end{equation}

To describe the particle-hole symmetry we first define operators
$\hat J_s^{(o)}$ and $\hat J_s^{(h)}$. The operator $\hat J_s^{(o)}$
 changes the sign of the wave-function each time there is a
spin $s$ electron on an odd site, while the operator $\hat J_s^{(h)}$ 
interchanges
 holes and particles for
a spin direction $s$.
\begin{equation}
\hat J_s^{(o)}=\prod_{j=0}^{n-1}(1-2\hat n_{2j+1,s})\quad 
\hat J_s^{(h)}=\prod_{j=1}^{2n}(c^\dagger_{js}+c_{js})
\label{J}
\end{equation}
It is convenient to introduce the following combinations of these operators:
\begin{equation}
\hat J^{(o)}=\hat J_\up^{(o)}\hat J_\dn^{(o)}\quad \hat J^{(h)}=
\hat J_\up^{(h)}
\hat J_\dn^{(h)}\quad
\hat I^{(o)}=\hat J^{(o)}\hat J^{(h)}\quad \hat Z_\up=
\hat J_\up^{(o)}\hat J_\dn^{(h)}
\label{ph}
\end{equation}
 Note the action of  
various symmetries on the creation and annihilation operators 
\begin{equation}
\hat Z_\up^\dagger c_{j\up}\hat Z_\up=(-1)^jc_{j\up}\quad 
\hat Z_\up^\dagger c_{j\dn}\hat Z_\up=-c^\dagger_{j\dn}
\label{Ztr}
\end{equation}
\begin{equation}
\hat I^{(o)}c_{js}\hat I^{(o)}=(-1)^{j+1}c^\dagger_{js}
\label{Iotr}
\end{equation}
\begin{equation}
\hat \sigma c_{js}\hat \sigma= c_{-j,s}
\label{sigmatr}
\end{equation}
\begin{equation}
\hat J^{(o)}c_{js}\hat J^{(o)}=(-1)^jc_{js}
\label{Jotr}
\end{equation}
It follows from equations (\ref{Ztr}) and (\ref{Iotr}) that
\begin{equation}
\hat Z_\up|m_\up, m_\dn\rangle=|m_\up, 2n-m_\dn\rangle
\label{dncons}
\end{equation}
and
\begin{equation}
\hat I^{(o)}|m_\up, m_\dn\rangle=|2n-m_\up, 2n-m_\dn\rangle
\label{totcons}
\end{equation}
where $|m_\up, m_\dn\rangle$ is a state with $m_\up$ spin up electrons
and $m_\dn$ spin down electrons, and $2n$ is the number of sites.
Thus, operators $\hat I^{(o)}$ and $\hat Z_\up$ conserve 
the number of particles 
only at a half
filling with an equal number of up and down electrons.

Equations (\ref{Ztr}) and (\ref{Iotr}) are used to evaluate various 
commutation relations for $\hat I^{(o)}$ and $\hat Z_\up$.
\begin{equation}
\{\hat Z_\up,\hat H\}=0\quad [\hat \sigma, \hat H]=0
\label{main}
\end{equation}
\begin{equation}
[\hat I^{(o)}, \hat H]=
\{\hat J^{(o)}, \hat T\}=\{\hat I^{(o)}, \hat S_z\}=
[\hat I^{(o)}, {\bf\hat S}^2]=0
\label{rest}
\end{equation}
\begin{equation}
\hat I^{(o)}\hat C_{2n}-\hat C_{2n}\hat I^{(o)}(-1)^{\bf\hat n}=
\hat Z_\up \hat C_{2n}+\hat C_{2n}\hat Z_\up(-1)^{{\bf \hat n}_\dn}=
\hat Z_\up\hat \sigma-(-1)^{n-1}\hat \sigma \hat Z_\up=0
\label{ambg}
\end{equation}
where $\{\hat A, \hat B\}\equiv \hat A\hat B+\hat B\hat A$, 
${\bf\hat n}={\bf\hat n}_\up+{\bf\hat n}_\dn$, and
$\hat T$
is the kinetic energy operator
$$
\hat T=T\sum_{j=1}^N\sum_{s=\up\dn}
(c^\dagger_{js}c_{j+1\,s}+c^\dagger_{j+1\,s}c_{js})
$$

 $\hat Z_\up$ is a unitary operator that 
anticommutes with the Hamiltonian. Therefore
it defines a mapping from the set of operators that commute with the 
Hamiltonian onto itself, i.e. $[\hat A, \hat H]=0$  implies
$[\hat Z_\up^\dagger\hat A \hat Z_\up, \hat H]=0$.
Using this transformation, we can construct a new $su(2)$ algebra
($\eta$-pairing $su(2)$) from the spin $su(2)$ (\ref{spinsu2}) 
\cite{su1, su2, su3, su4}.  
\begin{equation} 
 \hat L_z=\frac{{\bf\hat n}_\up+{\bf\hat n}_\dn}{2}-n\quad 
 \hat L_-=\sum_{j=1}^{2n}(-1)^jc_{j\up}c_{j\dn}\quad 
\hat L_+= (\hat L_-)^\dagger
\label{sunew}
\end{equation}
The Casimir operator of this new $su(2)$ 
\begin{equation} 
{\bf\hat L}^2=\hat Z^\dagger_\up{\bf\hat S}^2\hat Z_\up 
\label{Shat}
\end{equation}
preserves the number of particles and therefore its eigenvalues 
$L(L+1)$ can be
included in the set of quantum numbers used to label the eigenstates.
The combination $SU(2)\times SU(2)/{\bf Z}_2$ yields the complete $SO(4)$
symmetry of the Hubbard Hamiltonian \cite{su1, su2, su3, su4}.

\section{Conserved Currents}

Now let us turn to the parameter dependent integrals of motion. In principle,
an infinite number of these integrals can be obtained by methods outlined
in \cite{S1,S2,S3}. Clearly, on a finite lattice only a finite number
of these integrals are independent (see Section~6 for more details).
  The general form of the
$r$th conserved current is 
$$
\hat I_r(U, T)=\sum_{k=0}^l U^k T^{l-k}\hat I_r^k
$$
where $\hat I_r^k$ are parameter independent operators. 
The U--independent part of $\hat I_r$ has a simple form  
\begin{equation}
\hat I_r(U=0)=T^l\hat I_r^0=\pm
(i)^{r+1}T^l\sum_{j=1}^N\sum_{s=\uparrow\downarrow}
 (c^\dagger_{j+r,s}c_{js}-(-1)^r c^\dagger_{js}c_{j+r,s})
\label{tn}
\end{equation}

As far as permanent degeneracies are concerned, an important feature of
the currents is that odd ($r=2k+1$) and even ($r=2k$) currents transform 
differently
 (equations (\ref{oddsigma})--(\ref{evenZ}))
 under
spatial reflections ($\hat \sigma$) and partial particle-hole transformation
($\hat Z_\up$). Odd currents are in
many respects similar to the Hamiltonian, while properties of even currents 
are essentially different.  

  First few nontrivial 
  currents  were derived explicitly in 
\cite{S2,Lu,Gr,GM}.
( See also \cite{GM} and \cite{W} for a discussion on the derivation
and the structure of the higher conserved currents.) To analyze
 the benzene example we
will need the explicit form of only the first two nontrivial 
integrals of motion. Since $\hat I_1\equiv \hat H$, these two are
\begin{equation}
 \hat I_2=-iT\sum_{j=1}^N\sum_{s=\uparrow\downarrow}
 (c^\dagger_{j+2s}c_{js}-c^\dagger_{js}c_{j+2s})
-iU\sum_{j=1}^N\sum_{s=\uparrow\downarrow}
 (c^\dagger_{j+1s}c_{js}-c^\dagger_{js}c_{j+1s})(\hat n_{j+1,-s}+\hat n_{j,-s}
-1)
\label{j2}
\end{equation}

\begin{equation}
\begin{array}{rcl}
\dis\hat I_3&=&\dis T^3\sum_{j=1}^N\sum_{s=\uparrow\downarrow}
\left(c^\dagger_{j+3s}c_{js}+c^\dagger_{js}c_{j+3s}\right)
+\dis T^2U\sum_{j=1}^N\sum_{s=\uparrow\downarrow}\left\{
\left(c^\dagger_{j+1s}c_{j-1s}+c^\dagger_{j-1s}c_{j+1s}\right)\right.\\
\\
&&\dis \left(\hat n_{j+1,-s}+\hat n_{j,-s}+\hat n_{j-1,-s}-\frac{3}{2}\right) 
 +\left(c^\dagger_{j+1s}c_{js}-c^\dagger_{js}c_{j+1s}\right)
\left(c^\dagger_{j,-s}c_{j-1,-s}-c^\dagger_{j-1,-s}c_{j,-s}\right)-\\
\\
&&\dis \left(\hat n_{j+1s}-\frac{1}{2}\right)\left.
\left(\hat n_{j,-s }-\frac{1}{2}\right)\right\}
+T^2U\sum_{j=1}^N
\left\{\left(c^\dagger_{j+1\uparrow}c_{j\uparrow}-
c^\dagger_{j\uparrow}c_{j+1\uparrow}\right)
\left(c^\dagger_{j+1\downarrow}c_{j\downarrow}-
c^\dagger_{j\downarrow}c_{j+1\downarrow}\right)\right. \\
\\
&&\dis -\left(\hat n_{j\uparrow}-\frac{1}{2}\right)
\left.\left(\hat n_{j\downarrow}-\frac{1}{2}\right)\right\}
-TU^2\sum_{j=1}^N\sum_{s=\uparrow\downarrow}
 \left(c^\dagger_{j+1s}c_{js}+c^\dagger_{js}c_{j+1s}\right)
\left(\hat n_{j+1,-s}-\frac{1}{2}\right)\left(\hat n_{j,-s }-\frac{1}{2}\right)\\
\\
&&\dis -U^3/4
\sum_{j=1}^N\left(\hat n_{j\uparrow}-\frac{1}{2}\right)
\left(\hat n_{j\downarrow}-\frac{1}{2}\right)\\
\end{array}
\label{j3}
\end{equation}
 
Odd currents have the same U--independent symmetry as the 
Hubbard Hamiltonian, i.e.
in all commutation relations of Section~2 the Hamiltonian can be replaced 
with any
other odd current. For even currents two of the commutation relations are 
different.
Namely, for odd currents one can show that
\begin{equation}
[\hat \sigma, \hat I_{2k+1}]=0
\label{oddsigma}
\end{equation}
\begin{equation}
\{\hat Z_\up, \hat I_{2k+1}\}=0
\label{oddZ}
\end{equation}  
while for even currents
\begin{equation}
\{\hat \sigma, \hat I_{2k}\} =0
\label{evensigma}
\end{equation}
 \begin{equation}
[\hat Z_\up,\hat I_{2k}]=0
\label{evenZ}
\end{equation}
In other words, equations 
(\ref{oddsigma}--\ref{evenZ}) mean that even and odd currents 
have
different $\hat \sigma$ and $\hat Z_\up$ parities. For $k=1$ equations
(\ref{oddsigma}--\ref{evenZ}) have been derived in \cite{Gr}. 
Note also that since 
both $\hat \sigma$
and $\hat Z_\up$ are unitary and $I_r$ is Hermitian, equations 
(\ref{evensigma})
and (\ref{oddZ}) imply
\begin{equation}
Tr\hat I_r=0
\label{tr}
\end{equation}
Another  property of conserved currents which we will use bellow is
\begin{equation}
\hat J^{(o)}\hat I_r(T,U)\hat J^{(o)}=(-1)^{r+1}\hat I_r(-T,U)
\label{Tsign}
\end{equation}
Equations (\ref{oddsigma})--(\ref{evenZ}) and (\ref{Tsign})
 for $I_2$ and $I_3$ can be  
verified by 
inspection. For higher currents they can be derived by establishing the 
transformation
properties of the transfer matrix obtained in \cite{S1, S2, S3}. 

  Finally, let us measure $U$, $T$ and all other energies in units of $U-4T$.
This is equivalent to the replacement
\begin{equation}
U-4T=1\quad U=u\quad T=(u-1)/4 
\label{UT}
\end{equation}
We see  from (\ref{Tsign}) that
the spectrum and the eigenfunctions of $\hat I_r$ for 
$T\le0$ and $T\ge0$ are related via a simple transformation and it is 
therefore sufficient
to consider $T\le0$.
Thus, we have to compute the spectra only for $0\le u\le 1$.

\section{Diagonalization}

In this section we outline a method that we use 
for detailed numerical study of 
the   spectra
 of the Hamiltonian and conserved currents.   
In subsequent sections we will
compare the computer generated results for benzene ($m=n=3$) and several
other values of $m$ and $n$ to general predictions based on symmetry.

 From the results of Section 2 it follows 
that the set of quantum numbers that label the eigenstates of  the Hamiltonian
and
integrals of motion is $\{P, \sigma,  I^{(o)}, S, L\}$. 
However, there are certain 
restrictions on this set since not all symmetries and integrals of motion 
mutually commute. The quantum number $I^{(o)}$ can be assigned only when 
$m=n$ and $\sigma$ can be used only when $P=0$ or $P=n$ (see (\ref{totcons})
and (\ref{sigmam})). Also, since 
$\hat \sigma$ 
changes $P$ to $2n-P$ without affecting the energy and other quantum 
numbers, the spectrum for $2n>P>n$ is the exact copy of that for
$n>P>0$. Therefore, it is sufficient to diagonalize the Hamiltonian and 
the currents for $P$ ranging from 0 to $n$. Finally, since even currents
anticommute with $\hat \sigma$, their eigenvalues cannot be specified whenever
$\sigma$ is specified.

To utilize the conservation of momentum, we introduce in the usual way
the momentum 
space creation and annihilation operators 
$d^{\dagger}_{ps}$ and $d_{ps}$
\begin{equation}
d^{\dagger}_{ps}=\frac{1}{\sqrt{2n}}\sum_{j=1}^{2n} e^{i\pi pj/n}
c^{\dagger}_{js}\quad d_{ps}=\frac{1}{\sqrt{2n}}\sum_{j=1}^{2n} e^{-i\pi pj/n}
c_{js}\label{d}
\end{equation}
where $p=0,1,\dots,2n-1$.

Next, we choose a basis for $2m$-electron wave functions with $m$ spin up
and $m$ spin down electrons.
\begin{equation}
|{\bf r}\rangle=|{\bf p; q}\rangle=
d^{\dagger}_{p_1\up}\dots d^{\dagger}_{p_m\up}
d^{\dagger}_{q_1\dn}\dots d^{\dagger}_{q_m\dn}|0\rangle
\label{basis}
\end{equation}
with the following ordering convention
\begin{equation}
\begin{array}{l}
p_1<p_2<\dots<p_m\\
q_1<q_2<\dots<q_m\\
\end{array}
\label{order}
\end{equation}
Components of $|{\bf r}\rangle$ that are not in the interval $[ 0, 2n-1 ]$
have to be reduced modulo $2n$ into this interval. If the components of a
vector are not in the order (\ref{order}) they have to be permuted
to obtain these ordering and the wave function should be multiplied
by $(-1)$ if the permutation is odd. The total momentum of the state is
\begin{equation}
P=\sum_{k=1}^{2m} {\bf r}_k\qquad \mbox{(modulo $2n$)}
\label{P}
\end{equation}

From (\ref{Ztr}--\ref{sigmatr}) and (\ref{d}) we 
identify the action of operators of 
U--independent symmetry on basic states (\ref{basis})
\begin{equation}
\hat \sigma|{\bf r}\rangle=|{\bf -r}\rangle\label{sigmam}
\end{equation}
\begin{equation}
 \hat I^{(o)}|{\bf r}\rangle=
(-1)^{2m+P}|{\bf -\overline{(r+ne)}}\rangle\label{Im}
\end{equation}
\begin{equation}
  \hat Z_\up|{\bf p};{\bf q}\rangle=
(-1)^{(m+Q)}|{\bf (p+ne)};{\bf -\overline{q}}\rangle\label{Zm}
\end{equation}
where $Q=\sum_{k=1}^m q_k$ is the total momentum of spin down electrons,
$\overline{\bf r}$ denotes taking the complement of the set 
${\cal R}\equiv\{r_k\}$ in
the set of integers from 0 to $2n-1$. Finally, 
${\bf +ne}$ stands for adding $n$ to
each component of a vector.  

The rotation operator $\hat C_{2n}$ commutes with the Hamiltonian,
conserved currents, $\hat I^{(o)}$, 
${\bf\hat S}^2$, and ${\bf\hat L}^2$. Therefore,
 these operators split into blocks corresponding to
different values of $P$. This significantly reduces the size of the matrices
to be diagonalized. 
To generate the blocks for each value of
$P$ we need to derive the matrix elements of the Hamiltonian, integrals of
motion, and symmetry operators in the basis (\ref{basis}).
Equations  (\ref{sigmam}) and (\ref{Im}) allow
 to write down  $\hat \sigma$, $\hat I^{(o)}$, and $\hat Z_\up$ 
in the matrix notation.
Matrix elements of  the
Hamiltonian, and the operators ${\bf\hat S}^2$, $\hat I_2$, and 
$\hat I_3$ are summarized in Appendix~B.  

Now we have all ingredients needed to produce a computer program 
(e.g. on {\it Mathematica}) that generates exact blocks
of the Hamiltonian, symmetry operators, and first two currents for any
value of the total momentum $P$. 

After the value of the total momentum $P$ is chosen, 
the program picks one of the remaining U--independent symmetries, 
say $\hat \sigma$,
and finds a unitary transformation that makes it diagonal. The columns of
the matrix of this transformation  are
the eigenvectors of $\hat \sigma$. Since we know the eigenvalues of 
$\hat \sigma$  exactly $(\pm 1)$, determining the eigenvectors
for each eigenvalue reduces to a set of linear equations which can be
solved analytically.

 In this way, step by step, we diagonalize all
parameter independent symmetries and split the Hamiltonian and conserved
currents into smaller blocks. If the system is not too large, this can be
done analytically using {\it Mathematica}. For example, the complete
Hamiltonian for benzene is a $400\times400$ matrix, while individual blocks
range in size from $1\times1$ to $16\times16$.

 Finally, when all U--independent
symmetry is exhausted, we get exact blocks of the Hamiltonian and conserved
currents with all U--independent symmetry quantum numbers assigned to each
block. Each block is then diagonalized individually at different values of
$u$ ranging from 0 to 1 (see (\ref{UT})). 
Often, the small size of a block allows for
analytical diagonalization.

\section{Permanent Degeneracies and Other Spectral Properties}

\begin{figure}[p]
\epsfysize=3.0in 
\epsfxsize=4.0in  
\centerline{\epsfbox{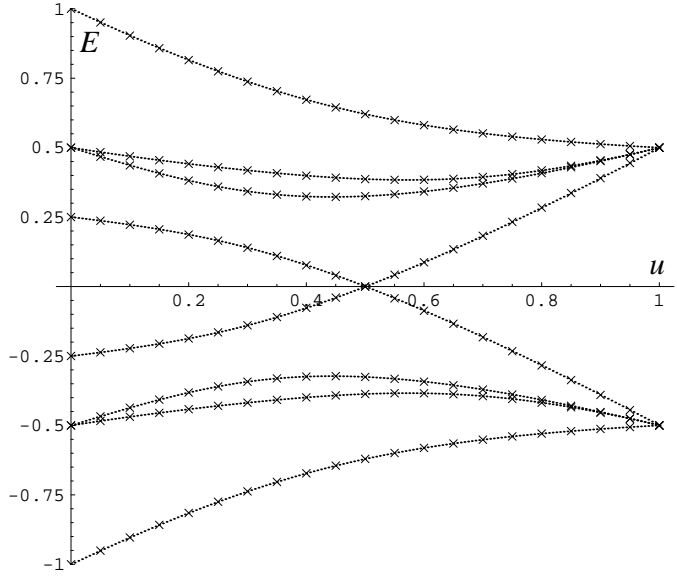}}
\caption{Energies for 
$ (P, \sigma, I^{(o)}, S, L )=
(0, 1, 1, 1, 1 )$ (cf. Fig~\ref{j31}) in units of $U-4T$ as functions
of $u=U/(U-4T)$. Levels
marked with crosses are twofold degenerate with respect to
$\sigma=\pm1$. They also correspond to 
$ (P, \sigma, I^{(o)}, S, L )=(0, -1, 1, 1, 1 )$ (equation~(\ref{I2psi})).
 Note that since 
$S=L$ the spectrum is symmetric under the reflection $E\to -E$ 
(equation~(\ref{Zpsi})).}
\label{h2}
\end{figure}

\begin{figure}[p]
\epsfysize=3.0in 
\epsfxsize=4.0in  
\centerline{\epsfbox{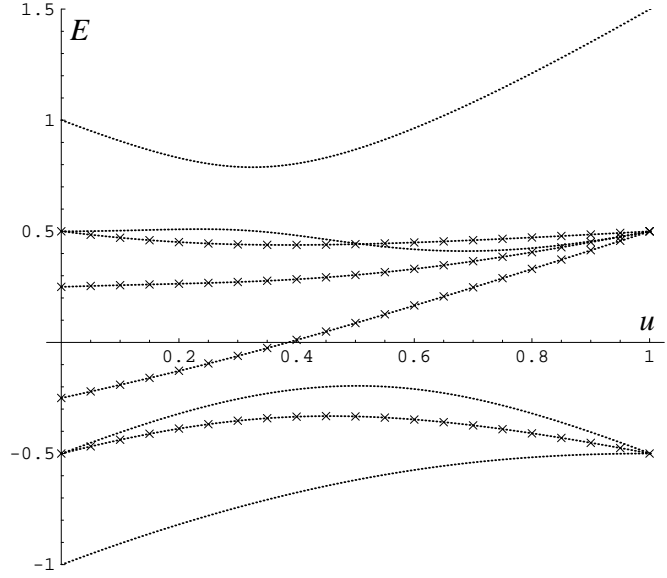}}
\caption{Energies for 
$ (P, \sigma,  I^{(o)}, S, L )=
(0, -1, -1, 0, 1 ) $ (cf.~Fig.~\ref{j32} and \ref{j21}) 
in units of $U-4T$ as functions
of $u=U/(U-4T)$. Levels
marked with crosses are twofold degenerate with respect to $\sigma=\pm1$.
 They also correspond to 
$ (P, \sigma, I^{(o)}, S, L )=(0, 1, -1, 0, 1 )$ (equation (\ref{I2psi})). 
Upside down the
figure shows levels for the same set of quantum numbers with 
$S \leftrightarrow L$ (equation (\ref{Zpsi})). Note the nondegenerate
 levels.}
\label{h3}
\end{figure}

As soon as the Hamiltonian (\ref{H}) is 
diagonalized, we discover numerous cases of permanent degeneracy, see
e.g. Table~\ref{nmbs} and Fig.~\ref{h1}, \ref{h2}, and \ref{h3}.
  We make the following observations. 
The majority of degeneracies are twofold degeneracies with respect
to $\sigma=\pm1$.  However,  there are also several 
multiply  
degenerate levels. The characteristic feature of these states is that 
 the energies are linear
in $U$.

\begin{table}[!ht]
\begin{center}
\begin{tabular}{|c|c|c|l|}
$P$ & $n$ & $m$ & (degree of degeneracy, number of levels)\\
\hline
0 &  3  & 2 & (1, 12); (2, 12)\\
\hline
3 &  3  & 2 & (1, 15); (2, 7); (3, 1); (7, 1)\\
\hline
0  &  3 & 3 & (1, 32); (2, 18)\\
\hline
3  &  3 & 3 & (1, 30); (2, 12); (7, 2)\\
\hline
0  &  4 & 2 &  (1, 28); (2, 31); (3, 1); (7, 1)\\
\hline
4  &  4 & 2 & (1, 15); (2, 41); (3, 1)\\
\hline
0  &  4 & 3 & (1, 70); (2, 150); (3, 2); (8, 2)\\
\hline
4  &  4 & 3 & (1, 56); (2, 155); (3, 1); (7, 1); (8, 2)\\
\hline
0  &  4 & 4 & (1, 126); (2, 229); (3, 1), (7, 1); (12, 2) \\
\hline
4  &  4 & 4 & (1, 126); (2, 229); (3, 1); (7, 1); (12, 2)\\
\end{tabular}
\caption{Degrees of degeneracy in the sectors 
$P=\{0, n\}$ for $n=\{3, 4\}$}
\label{nmbs}
 \end{center}
\end{table}

First, we analyze the twofold degeneracies. 
Let
$\psi(E, I_{2k+1}, \sigma,   S, L)$,
where $E$ is the energy, be an eigenstate of the Hamiltonian as well
as all other {\it odd}
currents and operators $\hat \sigma$, ${\bf \hat S}^2$, and ${\bf \hat L}^2$. 
Being an eigenstate of $\hat \sigma$, the state $\psi$ cannot be an
eigenstate of any even current since $\hat I_{2k}$ and $\hat \sigma$
anticommute. Let $\phi(E, I_{2k}, P, S, L, I_{2k+1})$ denote
an eigenstate of all even and
odd currents. On the other hand $ \phi\equiv\psi$ when $\sigma$ is 
not assigned, 
i.e. when $P\ne\{0, n\}$.
We note that
equation (\ref{evensigma}) means that either the state $\psi$ is
annihilated by the operator $\hat I_{2k}$, $\hat I_{2k}\psi=0$ for all $U$,
or
\begin{equation}
\hat I_{2k}
\psi(E, I_{2k+1}, \sigma, S, L)=
\psi(E, I_{2k+1}, -\sigma, S, L) 
\label{I2psi}
\end{equation}
Equations (\ref{evenZ}) and (\ref{evensigma}) imply 
\begin{equation}
\hat Z_\up\psi_n(E, I_{2k+1}, S, L)=
\psi_n(-E, -I_{2k+1}, L,  S)
\label{Zpsi}
\end{equation}
\begin{equation}
\hat Z_\up\phi_n(E, I_{2k},  S, L, I_{2k+1})=
\phi_n(-E, I_{2k}, L, S,  -I_{2k+1})
\label{Zphi}
\end{equation}
\begin{equation}
\hat \sigma\phi_n(E, I_{2k}, P, S, L, I_{2k+1})=
\phi_n(E, -I_{2k}, -P, S, L,  I_{2k+1})
\label{sigmaphi}
\end{equation}
It follows from equation (\ref{Zpsi}) that  for each state of 
energy $E$
and eigenvalues of odd currents $\{I_{2k+1}\}$ there is a state with the 
energy $-E$ and eigenvalues $\{-I_{2k+1}\}$. 
According to equation (\ref{dncons}) these two states have the same 
number of particles only if $m=n$. Equation (\ref{Zphi}) implies
that eigenstates of $I_{2k}$ are doubly degenerate. Finally, from (\ref{I2psi})
we conclude that  any state in the sector $P=\{0, n\}$
that is not annihilated by {\it all} $\hat I_{2k}$ is at least
twofold permanently degenerate. This applies to all values of $m$ and $n$.
In other words, all energy levels in 
the sector $P=\{0, n\}$ that are not
in the kernel of $\hat I_{2k}$ for all $U$ and $k$ are doubly degenerate
(see also \cite{Gr}).
 Consequently, all nondegenerate
states should be annihilated by any even current.

A twofold degeneracy is explained on symmetry grounds as soon as we
identify an even current that maps the two degenerate states into each other.
Unfortunately, we do not yet know how to prove for arbitrary $m$ and
$n$ that for any twofold
degeneracy there exists such an even current.  
 For benzene we checked numerically that all doubly 
degenerate states are maped into each other by $\hat I_2$ and 
therefore no other
even currents are needed to explain twofold degeneracies.

Since nondegenerate states are annihilated by any even current,
nondegenerate states exist only if all even currents have nontrivial kernels. 
Let us show that this necessary condition is met when $m$ is odd or $m\ne n$. 
  Indeed,
since $\hat I_{2k}$ maps the subspace $\sigma=+1$ into the subspace $\sigma=-1$
(\ref{I2psi}), its kernel has to contain at least $|d_{+1}-d_{-1}|$ 
states, where
$d_{\pm1}$ are the dimensions of $\sigma=\pm1$ subspaces. 
Since the eigenstates of
$\sigma=\pm1$ are $|{\bf r}\rangle\pm|{-\bf r}\rangle$ (\ref{sigmam}), 
$d_{+1}-d_{-1}$ is the number of states such that 
$|{\bf r}\rangle=|{-\bf r}\rangle$ minus the number of states such
that $|{\bf r}\rangle=-|{-\bf r}\rangle$. This is calculated to be 
\begin{equation}
d_{+1}-d_{-1}=\left\{{n-1\choose [\frac{m}{2}]}+(-1)^{m+1}
{n-1\choose [\frac{m-1}{2}]}\right\}^2
\label{dd}
\end{equation}
where $[x]$ is the integer part of $x$\footnote{The high degeneracy of the
zero eigenvalue of even integrals of motion suggests that
even conserved currents have an additional symmetry.
An interesting open question is whether the corresponding eigenstates can
be chosen to be independent of $U$ (compare to the discussion on multiply
degenerate levels of the Hamiltonian bellow).}.

The evaluation of an 
upper bound on the fraction of the nondegenerate states ($f$)
in the sector  $P=\{0, n\}$ can be reduced to a combinatorial problem.
We note that the number of such states  cannot exceed
 the dimension of  $kerI_2$ at $U=0$ restricted to the
subspace $P=\{0, n\}$. Let us denote this dimension by $K$.
 The kernel of $I_2$ at $U=0$ 
consists of all states
 $|{\bf r}\rangle$ such that
\begin{equation}
\sum_{k=1}^m\biggr[\sin\frac{2\pi p_k}{n}+
\sin\frac{2\pi q_k}{n}\biggl]=0
\label{ker}
\end{equation}
Thus we have $f\le K/D\equiv g$, where $D$ is the dimension of $P=\{0, n\}$
subspace. First, one
can show by explicitly constructing a $P=0$ 
state such that the left hand side of (\ref{ker}) is nonzero 
 that $K<D$ for $2n-2\ge m\ge 2$ and $n\ne 2$. This means that
twofold degenerate states exist for all values of $m$ and $n$ except $n=2$
and $m=\{0, 1, 2n-1, 2n\}$.
 Numerical values of $g$ for the first few 
 $m$ and $n$ are tabulated in Table~\ref{d/D}.
 Since $g(m, n)=g(2n-m, n)$, only values for $m\le n$ are
shown.

\begin{table}[!ht]
\begin{center}
\begin{tabular}{|c|c|c|c|c|c|}
$m$ \verb|\| $n$ & 2& 3& 4& 5 & 6\\
\hline
2  & 1.0 &0.89 &0.79 & 0.76 & 0.71\\
\hline
3  & -- & 0.88 & 0.765 & 0.70& 0.64\\
\hline
4  & -- & -- & 0.764 & 0.675 & 0.62\\
\hline
5  & -- &  -- & -- &  0.670 & 0.610\\
\hline
6  & -- & -- & -- &  -- &   0.608\\
\end{tabular}
\end{center}
\caption{Values of $K/D$. Note that the upper bound on the number of
nondegenerate states monotonically decreases as $m$ and $n$ increase.}
\label{d/D}
\end{table}
 
We see from Table~\ref{nmbs} 
that at small $m$ and $n$ a substantial fraction of levels are
nondegenerate. 
Next, we consider the limit $n\gg m$.
The exact value of $D$ depends on the greatest common
divider of $m$ and $n$. However, this  is not essential
for large $n$, since we can always decrease or increase $n$ by a small
number so that
$m$ and $n$ become mutually prime, in which case 
 the dimension of $P=\{0, n\}$ subspace is $1/n$
of the total dimension of the Hilbert space.
 \begin{equation}
D=\frac{1}{n}{2n\choose m}^2\sim n^{2m-1}\quad\mbox{for $n\gg m$}
\label{D}
\end{equation}
 To analyze the behavior
of $K$ in the limit $n\gg m$, we note that 
there are two ways how a number of terms in (\ref{ker})
can cancel. The first option is the pairwise cancellation. Alternatively,
$r$ terms can cancel if their momenta correspond to the vertices of 
the  regular $r$--gon. This can  
happen only if $n$ is divisible by $r$.  
For a given $p$ there are  at most
four other momenta with which it can cancel pairwise.
Therefore, for $n\gg m$ the number of ways in which
the sum (\ref{ker}) can cancel pairwise grows slower than $n^m$, while
the number of non-pairwise cancellations grows  slower than $n^{m-1}$. We 
conclude that
\begin{equation}
f\le \frac{\mbox{const}}{n^{m-1}}\quad\mbox{for $n\gg m$}
\label{lD}
\end{equation}
This asymptotics combined with numerical values from Table~\ref{d/D} suggests
that almost all states 
in the sector $P=\{0, n\}$ are at least twofold degenerate
in the thermodynamical limit.

As for the multiply degenerate states the first observation is that,
   at least for benzene, all of their eigenvalues are linear in the
coupling parameter $u$. This seems to suggest that these states are
 simultaneous eigenstates of the kinetic ($\hat T$)
and the potential ($\hat U$) energy operators (\ref{UT}).
 Indeed, this turns out to be the case for the majority of the multiply
degenerate levels for benzene.  However, the benzene Hamiltonian also has
 two eigenstates with $E=\pm u/2$ and $P=3$ that are not simultaneous
eigenstates of $\hat T$ and $\hat U$.  

Some of the multiply degenerate
states can be obtained using the prescription of \cite{su2}. The idea is
the following. We take a state $|{\bf r}\rangle$ that has only one species
of electrons, say $2l$ ($l\le n$) spin down electrons, and no spin up electrons
\begin{equation}
|{\bf r}\rangle=|0; {\bf q}\rangle=
|0;
 q_1, \dots q_{2l}\rangle
\label{down}
\end{equation}
Clearly, $|{\bf r}\rangle$ is an eigenstate of both $\hat T$ and $\hat U$.
\begin{equation}
\hat T|{\bf r}\rangle=\frac{u-1}{2}
\sum_{k=1}^{2l}\cos\frac{\pi q_k}{n}|{\bf r}\rangle\quad
\hat U|{\bf r}\rangle=u(n/2-l)|{\bf r}\rangle
\label{TU}
\end{equation}
On the other hand, $|{\bf r}\rangle$ is 
a lowest weight state in the $so(4)$ algebra 
constructed from ${\bf S}$ and ${\bf L}$
\begin{equation}
\begin{array}{l}
S_z|{\bf r}\rangle=-l|{\bf r}\rangle\quad S=l\\
\\
\hat L_z|{\bf r}\rangle=(l-n)|{\bf r}\rangle\quad L=n-l\\
\end{array}
\label{waites}
\end{equation}
The energy according to (\ref{TU}) is
\begin{equation}
E(u)=\frac{u-1}{2}\sum_{k=1}^{2l}\cos\frac{\pi q_k}{n}+u(n/2-l)
\label{mltpl}
\end{equation}
If $m\ge l$, we can get an eigenstate $|{\bf\tilde r}\rangle$  with
$m$ spin up and $m$ spin down electrons, by applying the raising
operators of the $so(4)$ algebra
\begin{equation}
 |{\bf\tilde r}\rangle=\hat S_+^l\hat L_+^{m-l}|{\bf r}\rangle
 \label{newst}
\end{equation}
The total number of states  that can be obtained in this way is
\begin{equation}
M=\sum_{l=0}^{m}{2n\choose 2l}
\label{total}
\end{equation}
while the number of states that have the same energy (\ref{mltpl}) cannot
exceed $\dis {2n\choose 2l}$. For $m=n=3$ the allowed choices for $l$ 
are $l=\{0, 1, 2, 3\}$. 
The choice $l=1$ yields 15 eigenstates of which 3 states
have the total momentum $P=3$ and are degenerate with the energy $E=u/2$.
By partial particle-hole symmetry (\ref{Zpsi})
 the choice $l=2$ also gives 3 states with
$P=3$ and the energy $E=-u/2$. The actual number of states of energy $u/2$ and
$P=3$ is 7 (Table~\ref{nmbs}). Wave functions of 6 of these
 states are independent
of $U$. 

We arrive at the conclusion that
 contrary to what was conjectured in \cite{su2}, not all
U--independent eigenstates can be obtained using the above prescription.
The six U--independent eigenstates together with the corresponding states
that have lowest weight in $so(4)$ are given in Appendix A.

Clearly, the multiple degeneracies discussed above mean that there is an
additional U--independent symmetry of the Hubbard Hamiltonian that have
been overlooked in Section 2. This symmetry seems to be related to the high
degeneracy of the spectra of the potential and kinetic energy operators.  
However, currently we do not know what is the algebra of generators
of this additional symmetry and how  its representations are to be classified.

\vspace{1cm}

\begin{figure}[p]
\epsfysize=3.0in 
\epsfxsize=4.0in  
\centerline{\epsfbox{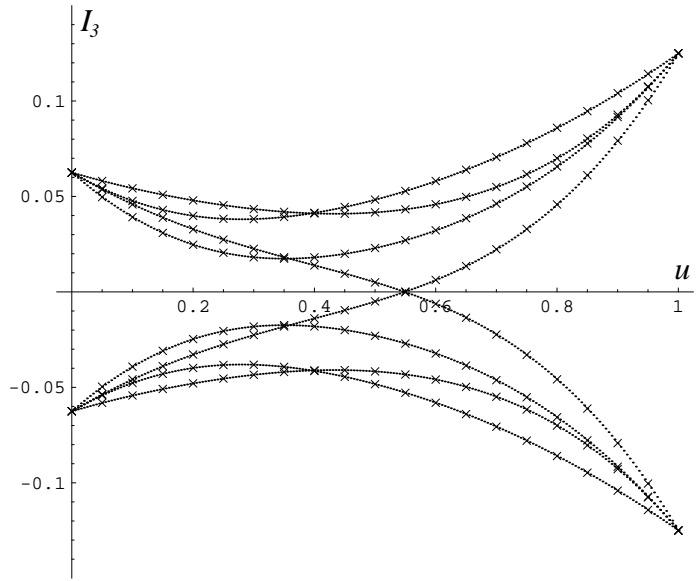}}
\caption{Eigenvalues of $\hat I_3$ for 
$ (P, \sigma, I^{(o)}, S, L )=
(0, 1, 1, 1, 1 )$ (cf.~Fig.~\ref{h2}) in units of $U-4T$ as functions
of $u=U/(U-4T)$. All eigenvalues are twofold degenerate with respect
to $\sigma=\pm1$.
 They also correspond to 
$ (P, \sigma, I^{(o)}, S, L )=(0, -1, 1, 1, 1 )$ (equation (\ref{I2psi})). 
The spectrum is symmetric under the
reflection $I_3\to -I_3$ (equation (\ref{Zpsi})). Note however that  
crossings of $\hat I_3$ and $\hat H$
 occur at different values of $u$ except $u=\{0,1\}$.}
\label{j31}
\end{figure}

\begin{figure}[p]
\epsfysize=3.0in 
\epsfxsize=4.0in  
\centerline{\epsfbox{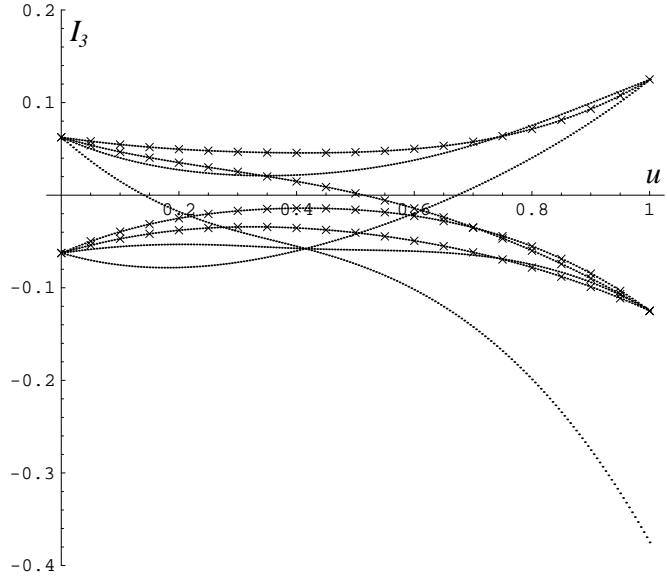}}
\caption{Eigenvalues of $\hat I_3$ for 
$ (P, \sigma, I^{(o)}, S, L )=
(0, -1, -1, 0, 1 )$ (cf.~Fig.~\ref{h3} and \ref{j21})
in units of $U-4T$ as functions
of $u=U/(U-4T)$. Levels
marked with crosses are twofold degenerate with respect to
$\sigma=\pm1$. They also correspond to 
$ (P, \sigma, I^{(o)}, S,L )=(0, 1, -1, 0, 1 )$ 
(equation (\ref{I2psi})). Upside down the
figure shows levels for the same set of quantum numbers with 
$S \leftrightarrow L$ (equation (\ref{Zpsi})).}
\label{j32}
\end{figure}

\begin{figure}[p]
\epsfysize=3.0in 
\epsfxsize=4.0in  
\centerline{\epsfbox{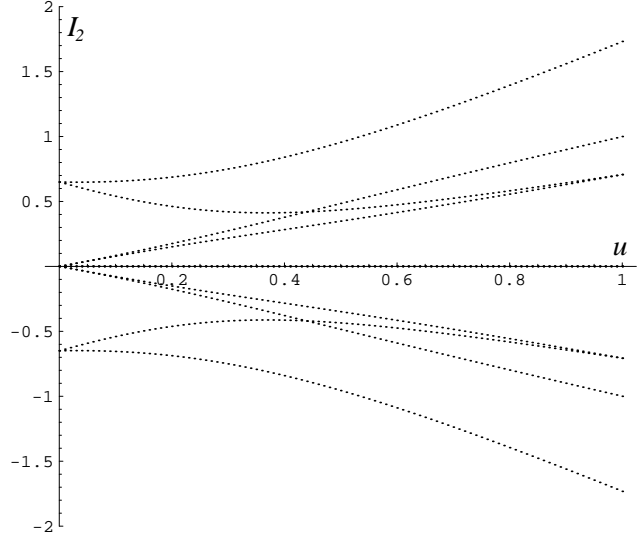}}
\caption{Eigenvalues of $\hat I_2$ for 
$ (P,   I^{(o)}, S, L )=
(0, -1, 0, 1 ) $ in units of $U-4T$ as functions
of $u=U/(U-4T)$. Note the difference as compared to Fig.~\ref{h3} and 
\ref{j32} The quantum number $\sigma$ cannot be assigned
since $\hat I_2$ anticommutes with $\hat\sigma$ (\ref{evensigma}).
 All eigenvalues are
twofold degenerate with respect to
$S \leftrightarrow L$ and $E\to -E$ (equation~(\ref{Zphi})). Since $P=-P=0$ 
the spectrum is symmetric under
the reflection (equation~(\ref{sigmaphi})).}
\label{j21}
\end{figure}

\begin{figure}[p]
\epsfysize=3.0in 
\epsfxsize=4.0in  
\centerline{\epsfbox{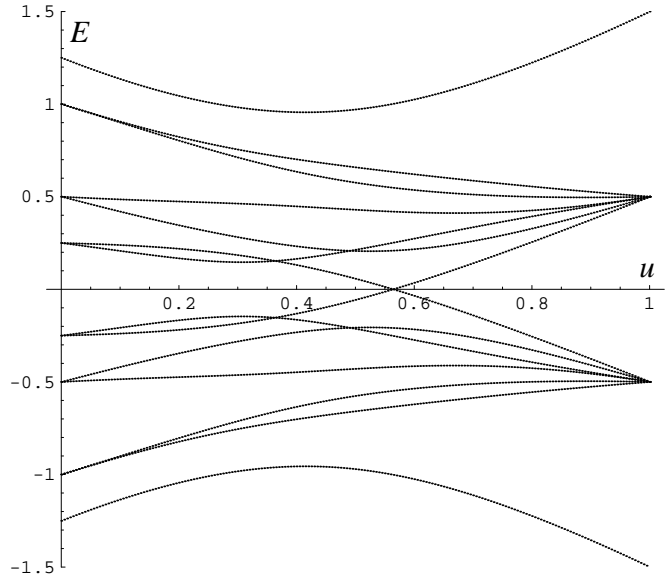}}
\caption{Energies for 
$ (P,   I^{(o)}, S, L )=
(2, 1, 0, 0 ) $ (cf.~Fig~\ref{j22} and \ref{j33}) 
in units of $U-4T$ as functions
of $u=U/(U-4T)$. Since $S=L$ the spectrum is symmetric under
the reflection $E\to -E$ (equation~(\ref{Zpsi})). Note the level crossings. }
\label{h4}
\end{figure}

\begin{figure}[p]
\epsfysize=3.0in 
\epsfxsize=4.0in  
\centerline{\epsfbox{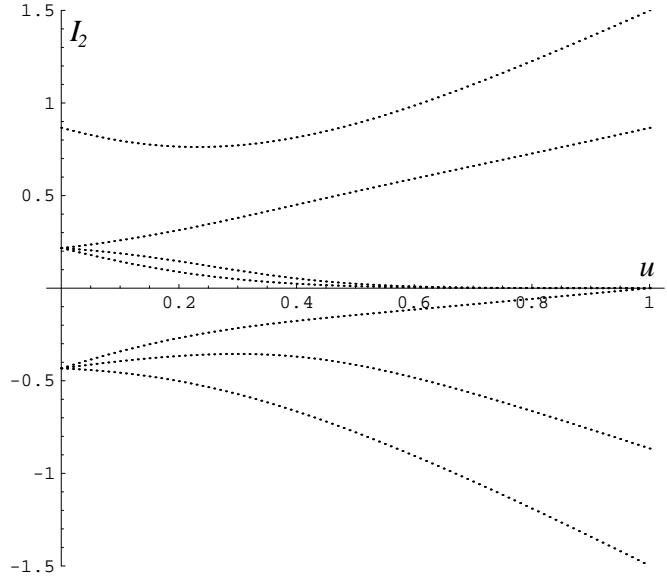}}
\caption{Eigenvalues of $\hat I_2$ for 
$ (P,   I^{(o)}, S, L )=
(2, 1, 0, 0 ) $ in units of $U-4T$ as functions
of $u=U/(U-4T)$. All eigenvalues are
twofold degenerate with respect to
$E \leftrightarrow  -E$ (equation~\ref{Zphi}). Note that unlike
Fig~\ref{h4} and \ref{j33} there is no reflection symmetry since
$P\ne -P$ (equation~(\ref{sigmaphi})).   }
\label{j22}
\end{figure}

\begin{figure}[p]
\epsfysize=3.0in 
\epsfxsize=4.0in  
\centerline{\epsfbox{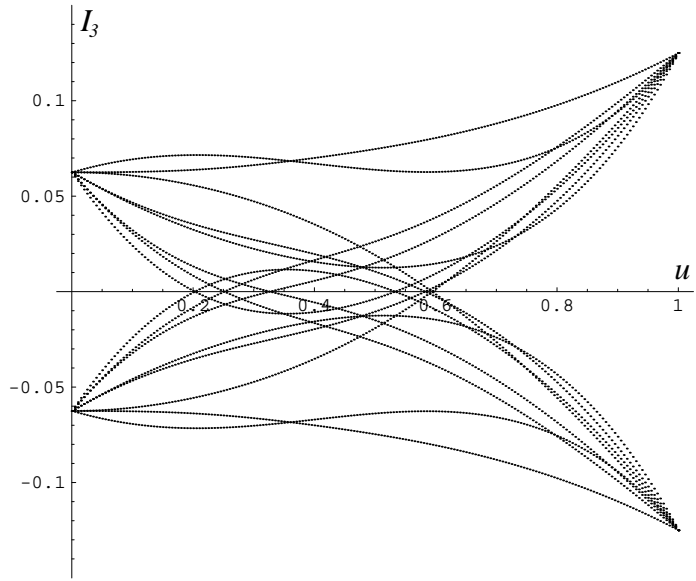}}
\caption{Eigenvalues of $\hat I_3$ for 
$ (P,   I^{(o)}, S, L )=
(2, 1, 1, 1 ) $ (cf.~Fig~\ref{h4} and \ref{j22})
 in units of $U-4T$ as functions
of $u=U/(U-4T)$. Since $S=L$ the spectrum is symmetric under
the reflection $E\to -E$ (equation (\ref{Zpsi})). }
\label{j33}
\end{figure}

\newpage 
\section{Level Crossings}

The spectra of the Hamiltonian and conserved currents display numerous cases
of crossings of levels that have the same U--independent symmetry quantum
numbers. In this section we argue that this type of behavior is  in
fact expected in integrable models. 

Let us first clarify the notion of a nontrivial integral of
motion for a quantum system on a finite lattice.
Let $H(u)=H_0+uV$ be a Hamiltonian that, in a certain basis,
 can be represented by an $s\times s$ matrix. For example, $H(u)$ can be one
of the blocks of the Hubbard Hamiltonian. Clearly, any matrix $M$ that is
an analytical function of $H(u)$ and $u$, $M=f(u, H(u))$, commutes with $H$.
Obviously, $M$ is not an independent conserved quantity.  
Therefore, we say that $J(u)$ is a nontrivial integral of
motion if $J(u)$ is hermitian, $[J(u), H(u)]=0$ and at the same time
 $J(u)$ cannot be written as an analytical
function of $H(u)$ and $u$. 

Now let us show that 
the existence of a nontrivial integral of motion implies
at least one level crossing.
Since $J(u)$ and $H(u)$ commute, there is a basis where both  these operators
are diagonal. Let $E_1(u),\dots, E_s(u)$ and $J_1(u),\dots, J_s(u)$ be
the eigenvalues of $H(u)$ and $J(u)$ respectively. Consider the
following set of algebraic
equations:
\begin{equation}
\begin{array}{lcl}
a_1 E_1^{s-1}(u)+\dots+a_{s-1}E_1(u)+a_s&=&J_1(u)\\
\quad\vdots&&\quad\vdots\\
a_1 E_s^{s-1}(u)+\dots+a_{s-1}E_1(u)+a_s&=&J_s(u)\\
\end{array}
\label{func}
\end{equation}
If equations (\ref{func}) have a solution, $J(u)$ and $H(u)$ are not
independent. Namely, $J(u)=a_1H^{s-1}+\dots+ a_s$. The system (\ref{func})
has no solutions if and only if $E_i(u^*)=E_j(u^*)$ and $J_i(u^*)\ne J_j(u^*)$
for some $i$, $j$, and $u^*$. Thus, the Hamiltonian can have a nontrivial
integral of motion only if it has a level crossing.

Let us analyze the suppression of level repulsion in integrable models in 
more detail.
Here we consider only real Hamiltonians that depend linearly on the
coupling constant $H(u)=H_0+uV$. 
We assume that there is a value of $u=\tilde u$ at which
all eigenvalues of $H(\tilde u)$ are nondegenerate. 
This is true for instance for blocks of the
Hubbard Hamiltonian for benzene. In general, since all integrals of motion
mutually commute, permanent degeneracies 
occur only with respect to   quantum numbers of
parameter independent symmetries.
Therefore, after all  these symmetries are factored out,
at a certain value of $u$  all states in a given block 
 are nondegenerate. Hence,
we can use the nondegenerate perturbation theory to write down the variation
of the eigenvalues  and the matrix elements of the perturbation with
$u$ in the vicinity of $\tilde u$. 
\begin{eqnarray}
\frac{d E_n(u)}{du}&=&V_{nn}(u)\label{E}\\
\frac{d V_{nn}(u)}{du}&=&2\sum_{j\ne n}\frac{V_{nj}^2(u)}{E_n(u)-E_j(u)}
\label{Vnn}\\
 \frac{d V_{nm}(u)}{du}&=&
\sum_{j\ne n}\frac{V_{nj}(u)V_{mj}(u)}{E_n(u)-E_j(u)}+
\sum_{j\ne m}\frac{V_{nj}(u)V_{mj}(u)}{E_m(u)-E_j(u)}
\label{Vnm}
\end{eqnarray} 
where  all matrix elements are evaluated in the running basis
$$
\begin{array}{l}
H\psi_n(u)=E_n(u)\psi_n(u) \\
V_{ij}(u)\equiv\langle \psi_i(u)|V|\psi_j(u)\rangle\\
\end{array}
$$

\begin{figure}[!p]
\epsfysize=3.0in 
\epsfxsize=4.0in  
\centerline{\epsfbox{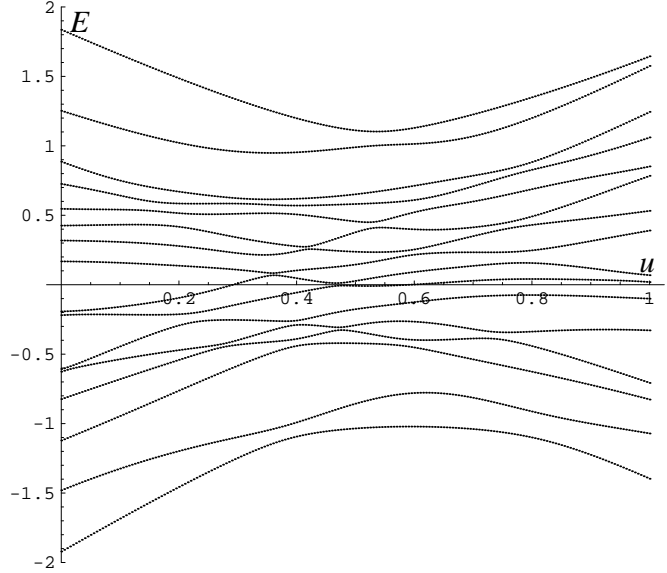}}
\caption{The block for 
$ (P,   I^{(o)}, S,L )=
(1, 1, 0, 1 ) $ has been perturbed by adding small random numbers to
$V_{ij}(\tilde u)$ and $E_i(\tilde u)$ at
$\tilde u=0.6$. Energies at other values of $u$ are determined according to
 (\ref{E})--(\ref{Vnm}) and plotted in units of $U-4T$ as functions
of $u=U/(U-4T)$.}
\label{pert}
\end{figure}

\begin{figure}[!p]
\epsfysize=3.0in 
\epsfxsize=4.0in  
\centerline{\epsfbox{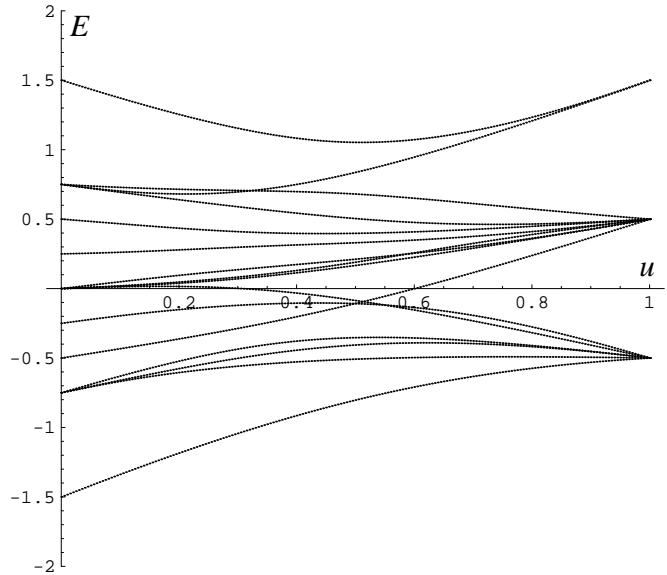}}
\caption{Unperturbed energies for 
$ (P,   I^{(o)}, S,L )=
(1, 1, 0, 1 ) $ in units of $U-4T$ as functions
of $u=U/(U-4T)$. }
\label{unpert}
\end{figure}

Equations (\ref{E})--(\ref{Vnm}) were used in \cite{Pechukas} to derive
the distribution of energy eigenvalues of the irregular spectrum in the
semiclassical limit.
Note  in particular that these equations  are
not model specific. The model itself is defined by the 
initial conditions --
the values of $V_{ij}(\tilde{u})$ and $E_i(\tilde{u})$. If
the initial conditions in any block are slightly perturbed\footnote{e.g.
by adding small random numbers to $E_i(\tilde{u})$ and $V_{ij}(\tilde{u})$.}, 
the integrability is lost and the majority of crossings are converted into
anticrossings (see e.g. Fig.~\ref{pert} and \ref{unpert}). 
At the same time, since the perturbation has the 
block diagonal structure of the original Hamiltonian, the perturbed Hamiltonian
still has the same U--independent symmetries.

Now consider a pair of levels that get close at $u=u_0$, i.e. we
assume that the absolute value of 
the energy difference between these two levels 
$\Delta(u)=E_1(u)-E_2(u)$
is much smaller 
than the energy distance to the remaining levels. Then, from 
(\ref{E}) and (\ref{Vnn}) we  obtain the following equations for   
$\Delta(u)$:
\begin{eqnarray}
\frac{d\Delta}{du}&=&V_{11}-V_{22}
\label{v}\\
\frac{d^2\Delta}{du}&=&\frac{4V^2_{12}}{\Delta}+F(u)
\label{F}
\end{eqnarray}
where
$$
F(u)= 2\sum_{j\ne 1,2}\biggr[\frac{V_{1j}^2(u)}{E_1(u)-E_j(u)}-
\frac{V_{2j}^2(u)}{E_2(u)-E_j(u)}\biggl]
$$


We can interpret $E_1$ and $E_2$ as  coordinates of two
one dimensional particles. Particles move with a relative
 velocity  $(V_{11}-V_{22})$ and interact with
a force $4V^2_{12}/\Delta+F(u)$.  
If  $V_{12}\lefteqn{\phantom{a}/}\to0$ as $\Delta\to0$, 
an infinite repulsion 
prevents the particles from colliding.   On the other hand, if levels 1 and 2 
have different 
U--independent symmetry, $V_{12}(u)$ is identically zero and levels 
are permitted to cross. These  arguments constitute 
the essence of the noncrossing rule. 

Let us incorporate conservation laws into the
picture. Let the corresponding block of the conserved current be 
$J(u^*)=J(u)+(u^*-u)W+(u^*-u)^2W'+\dots$ 
Since $J(u^*)$ commutes with $H(u^*)$ to all orders in $(u-u^*)$, we
get
\begin{equation}
[H(u^*), J(u^*)]=0\Rightarrow [H(u), W]+[V, J(u)]=0
\label{lin}
\end{equation}
Evaluating the matrix element of the above equation between states 1 and 2,
we find
\begin{equation}
V_{12}(u)=W_{12}(u)\frac{E_1(u)-E_2(u)}{J_1(u)-J_2(u)}
\label{V12}
\end{equation}
Therefore, unless $J_1(u)=J_2(u)$ in the vicinity of $u=u_0$, 
$V_{12}\approx\mbox{const}\,\Delta\to0$ as $\Delta\to0$ and levels are allowed
to cross. 

We have checked for benzene
that the eigenvalues of the Hamiltonian, $\hat I_2$, and $\hat I_3$ 
are never degenerate at the
same value of $u$ except at two points of special symmetry $u=0$ and $u=1$.
Because 
$\hat I_2$ also connects all doubly degenerate states,  we
conclude that $\hat I_2$  and $\hat I_3$ together with U--independent
symmetries explain all cases of level crossings and permanent degeneracies
in the spectrum of the Hubbard Hamiltonian for benzene.

In the  vicinity of  a crossing at $u=u^*$ the transverse
matrix element $V_{12}(u)$ can be expanded in series in the energy difference
$\Delta(u)$.
\begin{equation}
V_{12}(\Delta)=a_1\Delta+a_2\Delta^2+\dots
\label{trans}
\end{equation}
 The coefficient at the linear term can be derived
using the degenerate perturbation theory.
\begin{equation}
a_1=\sum_{j\ne 1,2} \frac{V_{1j}(u^*)V_{2j}(u^*)}{E_1(u^*)-E_j(u^*)}
\label{cc}
\end{equation}
Here the basis at $u=u^*$ has to be chosen so that $V_{12}(u^*)=0$ (see e.g.
Chapter 4 in \cite{LL}). 

The surprising feature of the Hubbard Hamiltonian is that, at least for 
benzene, the linear term
vanishes for all crossings at $u\ne\{0, 1\}$, 
while individual terms
in the summation (\ref{cc}) do not.  At this point, we do not have an 
explanation for this seemingly puzzling phenomenon. 

To see that this is not
a mere consequence of integrability, we study a simple example of
$3\times3$ Hamiltonians. This example also illustrates the connection
between crossings and integrability and is therefore interesting on its own
right.  

Let
\begin{equation}
H=H_0+uV\quad I=I_0+uW  
\label{int}
\end{equation}
be two $3\times3$ real symmetric matrices. We call a pair $(H, I)$ integrable
if these two matrices commute
\begin{equation}
[H, I]=0
\label{com}
\end{equation}
 Since
we can always add multiples of the identity matrix to $H$ and $I$ without
affecting integrability (\ref{com})
 or any of the level crossings, with no loss of
generality, we can assume that both $H$ and $I$ are traceless
$$
Tr H=Tr I=0
$$
Further, we say that a pair
$(H, I)$ is trivial if the $u$ dependence can be eliminated from either $H$
or $I$ by a $u$--independent unitary transformation (change of basis)
 and (or) by taking linear
combinations of $H$ and $I$. 

First we demonstrate that if $(H, I)$ is a nontrivial integrable pair, both
$H(u)$ and $I(u)$ have a single level crossing. 
Indeed, one can check that by changing the basis and taking linear
combinations
any nontrivial integrable 
pair can be brought to the following ``canonical'' form:
\begin{equation}
H=\left[\begin{array}{ccc}
1&0&0\\
0&1&0\\
0&0&-2\\
\end{array}\right]+u\left[
\begin{array}{ccc}
v_{11}&0&v_{13}\\
0&v_{22}&v_{23}\\
v_{13}&v_{23}&-v_{11}-v_{22}\\
\end{array}\right]
\label{3H}
\end{equation}
\begin{equation}
\phantom{a}I=\left[\begin{array}{ccc}
1&0&0\\
0&-1&0\\
0&0&0\\
\end{array}\right]+u\left[
\begin{array}{ccc}
w_{11}&w_{12}&w_{13}\\
w_{12} &w_{22}&w_{23}\\
w_{13}&w_{23}&-w_{11}-w_{22}\\
\end{array}\right]
\label{3I}
\end{equation}
Matrix elements of $W$ can be written in terms of $v_{ij}$'s using (\ref{com}).
\begin{eqnarray}
w_{11}&=&\frac{5v_{11}^2-v_{11}v_{22}+2v_{13}^2-4v_{22}^2-4v_{23}^2}{9(v_{11}-
v_{22})}\label{w11}\\
w_{12}&=&\frac{2v_{13}v_{23}}{3(v_{11}-v_{22})}\quad w_{13}=v_{13}/3
\label{w12}\\  
w_{22}&=&-\frac{4v_{11}^2+4v_{13}^2+
v_{11}v_{22}-5v_{22}^2-2v_{23}^2}{9(v_{11}-
v_{22})}\label{w22}\\
w_{23}&=&-v_{23}/3\quad w_{33}=-w_{11}-w_{22}\label{w33}
\end{eqnarray}
Since $W$ has five independent matrix elements, while $V$ has only four,
$w_{ij}$'s are constrained by
\begin{equation} 
w_{33}+\frac{w_{13}}{3w_{23}}+\frac{w_{23}}{3w_{13}}-
\frac{2}{3}w_{13}w_{23}=0
\label{constr} 
\end{equation} 

$H(u)$ has an apparent crossing at $u=0$. Further, it turns out 
that the
matrix $I(u)$ has a crossing if and only if 
the constraint (\ref{constr}) is met! This can be verified either directly
or using \cite{slava}, where, among other things, 
degeneracies of real symmetric $3\times3$
matrices are analyzed in detail. The crossing of $I$ is at
\begin{equation}
2/u^*=(w_{22}- w_{11}) + w_{12}(w_{13}/w_{23} - w_{23}/w_{13})
\label{u*}
\end{equation}
Similarly, it can be shown that linear combinations of $H$ and $I$ always
have crossings too. Therefore, if $(H, I)$ is a nontrivial integrable
pair, both $H(u)$ and $I(u)$ have a single level crossing.

Now let us show that the converse is also true, i.e. that
any real symmetric $3\times3$ matrix that has a pairwise crossing is also
integrable. We start with a
 real symmetric $3\times3$ matrix 
$H=H_0+uV$ that has a pairwise
crossing at $u=u^*$. By redefining the parameter $u\to u-u^*$ and a suitable
choice of basis we can write $H$ in the form (\ref{3H}). Therefore, there
exists a matrix $I$ of the form (\ref{3I}) such that $[I, H]=0$. In view of 
this 
explicit construction of  a ``dynamical conservation law'',
 we may say that  there are no accidental degeneracies in 
the case of $3\times 3$ real matrices.

Finally, we note that the linear term in 
expansion (\ref{trans}) is non zero for both $H$ and $I$
unless specially arranged.

\section{Conclusion}

\subsection{Summary}

We have shown that for all $m$ and $n$ except $m=\{0, 1, 2n-1, 2n\}$ and
$n=2$ there are twofold
permanent degeneracies for the values of the total momentum $P=0$ and $P=n$.
These degeneracies are a consequence of different transformation properties of
even and odd currents with respect to spatial reflections ($\hat\sigma$) 
and partial particle-hole transformation ($\hat Z_\up$) 
(equations (\ref{oddsigma}--\ref{evenZ})). We have argued that in the
thermodynamical limit the fraction of doubly degenerate states in the
sector $P=\{0, n\}$ approaches one. For benzene ($m=n=3$) we have checked
that all doubly
degenerate states are maped into each other by the first even current 
$\hat I_2$. We have also seen that some states can be nondegenerate since all
even currents have nontrivial kernels.

We have learned from the benzene example that there is a number of 
multiply degenerate states that are simultaneous
eigenstates of the potential and kinetic energy operators. The wave functions
of these states do not depend on the coupling $U$. This suggests the
existence of an additional U--independent symmetry. However, as we have
seen in Section~5 and Appendix~A some of these states cannot be obtained by
applying the raising (lowering) operators of the $so(4)$ algebra to
states that have only one species of electrons. 

Further, we have demonstrated how the constraints of the noncrossing rule are
lifted in integrable models. In fact, we have argued that in integrable
models crossings of levels with the same parameter
independent symmetry quantum numbers are expected. 
This was also illustrated by a simple example of $3\times 3$ matrix 
models where there is a one to one correspondence between crossings and
integrability. We have seen that for $m=n=3$ the first two nontrivial conserved
currents are sufficient to explain all cases of level crossings. We have
shown numerically 
that in the case of benzene the transverse matrix element is quadratic
in energy difference in the vicinity of all pairwise level crossings.

\subsection{Open Questions}

\begin{enumerate}

\item What is the symmetry responsible for the multiply (more than twofold)
 degenerate levels in the
spectrum of the Hubbard Hamiltonian?  Are energies of  multiply
degenerate states always linear in the coupling parameter? 
How one can count the
degrees of degeneracy and the number of multiply degenerate states
for arbitrary $m$ and $n$?

\item What is the symmetry responsible for the high degeneracy of the zero
eigenvalue of even integrals of motion?

\item For given $m$ and $n$ what is the exact number of states such that
 $\hat I_{2k}\psi=0$ for all $k$ and $U$? What is the exact number of
twofold degenerate states with $P=\{0, n\}$?

\item Is  the transverse matrix element always quadratic
in energy difference in the vicinity of all pairwise level crossings
in the Hubbard model? What is the explanation of this phenomenon?

\end{enumerate}

\section{Appendix A}

Here we write down  U--independent eigenstates that have $E=u/2$. 
See Section 4 for notations.

\begin{equation}
\begin{array}{lcl}
|{\bf r}_1\rangle&=&\dis-
|0, 1, 2; 1, 2, 3\rangle-1/2|0, 1, 3; 0, 2, 3\rangle
 +  1/2|0, 3, 4; 0, 3, 5\rangle+\\
\\
&&\dis |0, 4, 5; 3, 4, 5\rangle- 1/2|1, 2, 4; 1, 2, 5\rangle
 +  1/2|1, 4, 5; 2, 4, 5\rangle+(\up\leftrightarrow\dn)\\
\end{array}\quad 
\begin{array}{cc}
S=0&L=2\\
S_z=0&L_z=0\\
\end{array}
\label{1}
\end{equation}
\vspace{2mm} 

\begin{equation}
\begin{array}{lcl}
|{\bf r}_2\rangle&=&\dis|0, 2, 5; 1, 3, 4\rangle+|0, 3, 4; 1, 2, 5\rangle
+|0, 3, 5; 1, 2, 4\rangle\\
\\
&&\dis  -|1, 4, 5; 0, 2, 3\rangle-|2, 3, 5; 0, 1, 4\rangle
-|2, 4, 5; 0, 1, 3\rangle+(\up\leftrightarrow\dn)\\
\end{array}
\quad 
\begin{array}{cc}
S=0&L=1\\
S_z=0&L_z=0\\
\end{array}
\label{2}
\end{equation}
\vspace{2mm}

\begin{equation}
\begin{array}{lcl}
|{\bf r}_3\rangle&=&\dis -|0, 2, 5; 1, 3, 4\rangle+ 1/2
|0, 3, 4; 0, 3, 5\rangle+ 1/2|1, 2, 4; 1, 2, 5\rangle\\
\\
&&\dis  1/2|1, 4, 5; 2, 4, 5\rangle+ 1/2|0, 1, 3; 0, 2, 3\rangle
+|0, 1, 4; 2, 3, 5\rangle-(\up\leftrightarrow\dn)\\
\end{array}\quad 
\begin{array}{cc}
S=1&L=2\\
S_z=0&L_z=0\\
\end{array}
\label{3}
\end{equation}
\vspace{2mm}

\begin{equation}
\begin{array}{lcl}
|{\bf r}_4\rangle&=&\dis -|0, 1, 3; 0, 2, 3\rangle -|0, 3, 4; 0, 3, 5\rangle+\\
\\
&&|1, 2, 4; 1, 2, 5\rangle+|1, 4, 5; 2, 4, 5\rangle+(\up\leftrightarrow\dn)\\
 \end{array}\quad 
\begin{array}{cc}
S=0&L=2\\
S_z=0&L_z=0\\
\end{array}
\label{4}
\end{equation}
\vspace{2mm}

\begin{equation}
\begin{array}{lcl}
|{\bf r}_5\rangle&=&\dis |0, 1, 2; 1, 2, 3\rangle+|0, 1, 4; 2, 3, 5\rangle
|0, 1, 5; 2, 3, 4\rangle+\\
\\
&&|0, 2, 4; 1, 3, 5\rangle+|0, 2, 5; 1, 3, 4\rangle+|0, 4, 5; 3, 4, 5\rangle
-(\up\leftrightarrow\dn)\\
\end{array}\quad 
\begin{array}{cc}
S=1&L=2\\
S_z=0&L_z=0\\
\end{array}
\label{5}
\end{equation}
\vspace{2mm}

\begin{equation}
\begin{array}{lcl}
|{\bf r}_6\rangle&=&\dis 1/2|0, 1, 3; 0, 2, 3\rangle +
|0, 1, 5; 2, 3, 4\rangle-|0, 2, 4; 1, 3, 5\rangle-\\
\\
&&\dis 1/2|0, 3, 4; 0, 3, 5\rangle- 1/2|1, 2, 4; 1, 2, 5\rangle
+ 1/2|1, 4, 5; 2, 4, 5\rangle
 -(\up\leftrightarrow\dn)\\
\end{array}\quad 
\begin{array}{cc}
S=1&L=2\\
S_z=0&L_z=0\\
\end{array} 
\label{6}
\end{equation}

The corresponding lowest weight eigenstates of the Hamiltonian are 

\begin{equation}
|\tilde{\bf r}_1\rangle=-|1; 2\rangle-|2; 1\rangle+|4; 5\rangle+
|5; 4\rangle\quad S_z=0\quad L_z=-2
\label{1a}
\end{equation}

\begin{equation}
\begin{array}{lcl}
|\tilde{\bf r}_2\rangle&=&|0, 1; 2, 3\rangle-|0, 2; 1,3\rangle
-2|0, 3; 1, 2\rangle-2|0, 3; 4, 5\rangle-|0, 4; 3, 5\rangle+
|0, 5; 3, 4\rangle\\
\\
&&-2|1, 2; 0, 3\rangle-|1, 3; 0, 2\rangle+
|2, 3; 0, 1\rangle+|3, 4; 0, 5\rangle-|3, 5; 0, 4\rangle-
2|4, 5; 0, 3\rangle
\end{array}\quad 
\begin{array}{l}
S_z=0\\
L_z=-1\\
\end{array} 
\label{2a}
\end{equation}

\begin{equation}
|\tilde{\bf r}_3\rangle=|;1, 2\rangle+|;4, 5\rangle\quad S_z=-1\quad
L_z=-2
\label{3a}
\end{equation}

\begin{equation}
|\tilde{\bf r}_4\rangle=-2|0; 3\rangle-2|3; 0\rangle+|1; 2\rangle
+|2; 1\rangle+|4; 5\rangle+|5; 4\rangle\quad S_z=0\quad L_z=-2
\label{4a}
\end{equation}

\begin{equation}
|\tilde{\bf r}_5\rangle=|;0, 3\rangle\quad S_z=-1\quad L_z=-2
\label{5a}
\end{equation}

\begin{equation}
|\tilde{\bf r}_6\rangle=|; 1, 2\rangle-|; 4, 5\rangle
\quad S_z=-1\quad L_z=-2
\label{6a}
\end{equation}

States $|\tilde{\bf r}_3\rangle$, $|\tilde{\bf r}_5\rangle$, and
$|\tilde{\bf r}_6\rangle$  describe one species of electrons (spin down).
The remaining states cannot be obtained by the prescription detailed in
the end of Section 5.

\section{Appendix B}

In this section we collect the expressions for the matrix elements of
the Hamiltonian, $\hat I_2$, $\hat I_3$ and ${\bf\hat S}^2$.
Let $\tilde{p}_1<\dots<\tilde{p}_t$ be
the momenta of spin up electrons that are in ${\cal P}=\{p_k\}$ but not in
${\cal P'}=\{p'_k\}$, $\tilde{p}_{t+1}<\dots<\tilde{p}_{2t}$ be the
momenta that are in ${\cal P'}$
but not in ${\cal P}$. Similarly, we can define ${\cal Q}=\{q_k\}$,  
$\tilde{q}_1<\dots<\tilde{q}_u$, and $\tilde{q}_{u+1}<\dots<\tilde{q}_{2u}$.
The number of elements in the intersection of ${\cal P}$ with ${\cal P}'$
and ${\cal Q}$ with ${\cal Q}'$ is
therefore $L({\cal P}\cap {\cal P}')=m-t$ and 
$L({\cal Q}\cap {\cal Q}')=m-u$.
We also define
\begin{equation}
\epsilon({\bf r}',{\bf r})=\langle  {\bf r}'|
d^\dagger_{\tilde{p}_{t+1}\up}d_{\tilde{p}_1\up}\dots
d^\dagger_{\tilde{p}_{2t}\up}d_{\tilde{p}_t\up}
d^\dagger_{\tilde{q}_{u+1}\dn}d_{\tilde{q}_1\dn}\dots
d^\dagger_{\tilde{q}_{2u}\dn}d_{\tilde{q}_u\dn}|{\bf r}\rangle=\pm1
\label{eps}
\end{equation}
In this notation the matrix elements of ${\bf S}^2$, $H$, $\hat I_2$ and
$\hat I_3$ take the following form:
\begin{equation}
\begin{array}{lcll}
\dis \langle {\bf r}'|{\bf S}^2|{\bf r}\rangle
         &=&\dis n-L({\cal P}\cap {\cal Q})&\quad 
                   \mbox{if $|{\bf r}\rangle=|{\bf r}'\rangle $}\\
         &=&\dis \epsilon({\bf r}',{\bf r})  &\quad
               \mbox{if $\cal R=R'$ and $u=t=1$ }\\
         &=& 0&\quad\mbox{otherwise}\\
\end{array}
\label{S2mtrx}
\end{equation}
\begin{equation}
\langle {\bf r}'|H|{\bf r}\rangle=
\frac{u-1}{2}\sum_{k=1}^{2m}\cos \frac{\pi r_k}{n}\delta_{{\bf r}',{\bf r}}
-\frac{u}{2n}\epsilon({\bf r}',{\bf r})\delta_{u1}\delta_{t1}\delta_{PP'}
\label{Hmtrx}
\end{equation}
\begin{equation}
\langle {\bf r}'|\hat I_2|{\bf r}\rangle= 
\frac{u-1}{4}\sum_{k=1}^{2m}\sin \frac{2\pi r_k}{n}\delta_{{\bf r}',{\bf r}}
-\frac{u}{2n}\epsilon({\bf r}',{\bf r})\delta_{u1}\delta_{t1}\delta_{PP'}
\label{I2mtrx}
\end{equation}
$$
\begin{array}{lcl}
\dis \langle {\bf r}'|\hat I_3|{\bf r}\rangle&=&
\dis \frac{(u-1)^3}{128}\sum_{k=1}^{2m}
\sin\frac{3\pi r_k}{n}\delta_{{\bf r}',{\bf r}}-\frac{u^2(u-1)}{8n^2}
\bigl[\sum_{k,j,l=1}^{m}\cos\frac{\pi p_k}{n}\cos\frac{\pi(q_j-q_l)}{n}
\delta_{{\bf r}',{\bf r}}
+p\lr q\bigr]-\\
\\
&&\dis \frac{u(u-1)^2}{8}\sum_{k=1}^{m}
\sin\frac{\pi p_k}{n}\cos\frac{\pi q_k}{n}\delta_{{\bf r}',{\bf r}}
+\epsilon({\bf r}',{\bf r})\frac{u^2(u-1)}{8n^2}\sum_{k=1}^{m}\Bigl[
\cos\frac{\pi(\tilde{p}_2+\tilde{q}_2-p'_k)}{n}+\\
\\
&&\dis \cos\frac{\pi(\tilde{p}_2-\tilde{q}_1-p'_k)}{n}+p\lr q\Bigr]\delta_{u1}
\delta_{t1}\delta_{PP'}+
\epsilon({\bf r}',{\bf r})\frac{u^2(u-1)}{4n^2}\Bigl[
(\cos\frac{\pi(\tilde{q}_4-\tilde{q}_2)}{n}-\\
\end{array}
$$
\begin{equation}
\begin{array}{lcl}
&&\dis \cos\frac{\pi(\tilde{q}_4-\tilde{q}_1)}{n}\Bigr]\delta_{PP'}
\sum_{k=1}^m\cos\frac{\pi p'_k}{n}\delta_{u2}\delta_{t0}+
(p\lr q, u\lr t)\bigr]+\\
\\
&&\dis \epsilon({\bf r}',{\bf r})\frac{u^2(u-1)}{8n^2}\biggr[\Bigr(
\cos \frac{\pi(\tilde{q}_4-\tilde{q}_2-\tilde{p}_1)}{n}+
\cos \frac{\pi(\tilde{q}_4-\tilde{q}_2-\tilde{p}_2)}{n}+
\cos \frac{\pi(\tilde{q}_3-\tilde{q}_2-\tilde{p}_1)}{n}+\\
\\
&&\dis\cos \frac{\pi(\tilde{q}_3-\tilde{q}_2+\tilde{p}_2)}{n}\Bigl)\delta_{u1}
\delta_{t2}+(p\lr q, u\lr t)\biggl]\delta_{PP'}-\\
\\
&&\dis\epsilon({\bf r}',{\bf r})
\frac{u(u-1)^2}{16n}\biggr[\cos\frac{\pi(\tilde{p}_1+\tilde{p}_2)}{n}
\bigr(2\cos\frac{\pi(\tilde{q}_1-\tilde{q}_2)}{n}+1\bigl)+
\cos\frac{\pi(\tilde{p}_1+\tilde{q}_2)}{n}+p\lr q-\\
\\
&&\dis \cos\frac{\pi(\tilde{p}_1-\tilde{q}_1)}{n}-
\cos\frac{\pi(\tilde{p}_2-\tilde{q}_2)}{n}-
\cos\frac{\pi(\tilde{p}_2-\tilde{p}_1)}{n}+
\cos\frac{\pi(\tilde{p}_1+\tilde{q}_1)}{n}-\\
\\
&&\dis
\cos\frac{\pi(\tilde{p}_2-\tilde{q}_1)}{n}-1/2\biggl]
\delta_{u1}\delta_{t1}\delta_{PP'}\\
\end{array}
\label{I3mtrx}
\end{equation}

\end{document}